\documentclass[12pt]{article}%
\usepackage{amssymb}
\usepackage{amsfonts}
\usepackage{sw20elba}
\usepackage{amsmath}
\usepackage{graphicx}%
\providecommand{\U}[1]{\protect\rule{.1in}{.1in}}

\begin{document}

\title{Studies of Superconductivity and Structure for CaC$_{6}$ to Pressures above 15 GPa}
\author{M. Debessai*, J. J. Hamlin**, and J. S. Schilling\\{\small Department of Physics, Washington University, CB 1105}\\{\small One Brookings Dr, Saint Louis, Missouri 63130}
\and Y. Meng\\{\small HPCAT, Carnegie Institution of Washington, Argonne, Illinois 60439}
\and D. Rosenmann***, D. G. Hinks, and H. Claus\\{\small Materials Science Division, Argonne National Laboratory, Argonne,
Illinois 60439}}
\date{April 20, 2009 }
\maketitle

\begin{abstract}
The dependence of the superconducting transition temperature $T_{c}$ of
CaC$_{6}$ has been determined as a function of hydrostatic pressure in both
helium-loaded gas and diamond-anvil cells to 0.6 and 32 GPa, respectively.
Following an initial increase at the rate +0.39(1) K/GPa, $T_{c}$ drops
abruptly from 15 K to 4 K at $\sim$ 10 GPa. Synchrotron x-ray measurements to
15 GPa point to a structural transition near 10 GPa from a rhombohedral to a
higher symmetry phase.\vspace{1.5in}

\noindent*Present address: \ Institute for Shock Physics, Washington State
University, Pullman, Washington 99164

\noindent**Present address: \ Department of Physics and Institute for Pure and
Applied Physical Sciences, University of California, San Diego, La Jolla,
California 92093

\noindent***Present address: \ Center for Nanoscale Materials, Argonne
National Laboratory, Argonne, Illinois 60439

\end{abstract}

\section{Introduction}

The $s,p$-electron metal CaC$_{6}$ possesses with $T_{c}\approx11.5$ K the
highest superconducting transition temperature of all known graphite-related
compounds \cite{weller,emery1}. Magnetic susceptibility measurements by two
separate groups have shown that $T_{c}$ increases under pressure at the rate
$\sim+0.50(5)$ K/GPa (to 1.2 GPa) \cite{smith} or +0.42 to +0.48 K/GPa (to 1.6
GPa) \cite{kim}, where the pressure medium used was, respectively, kerosene or
silicon oil. This relatively large positive value of the initial slope
$dT_{c}/dP$ stands in contrast to the negative dependence normally found in
$s,p$-electron metals which superconduct at ambient pressure \cite{schilling1}%
. An electrical resistivity study on CaC$_{6}$ to 16 GPa using Fluorinert
pressure medium reported that $T_{c}$ initially increases under pressure at
the rate $\sim+0.5$ K/GPa, reaching a maximum value of 15.1 K before dropping
abruptly at 9 GPa to $\sim$ 5 K, apparently due to a first-order structural
phase transition \cite{gauzzi}. Recent room-temperature synchrotron x-ray
diffraction studies to 13 GPa using helium or argon pressure media report the
onset of reversible peak-broadening above 9 GPa which is interpreted as giving
evidence for an unusual order-disorder phase tranformation with no change in
space group symmetry from $R\overline{3}m$ rhombohedral \cite{gauzzi2}.
Further $T_{c}(P) $ studies to higher more nearly hydrostatic pressures would
clearly be useful to establish whether or not the abrupt drop in $T_{c}$ near
9 GPa is an intrinsic effect and to more fully characterize possible
structural transitions in this pressure region. Dense He is the most nearly
hydrostatic pressure medium available in experiments at low temperatures.

In the present work the pressure dependence of $T_{c}$ for CaC$_{6}$ is
studied in ac susceptibility measurements both in a He-gas pressure system to
0.6 GPa and in a diamond-anvil-cell (DAC) to 32 GPa utilizing dense He
pressure medium. Following an initial increase under pressure to 15 K at the
rate +0.39(1) K/GPa, a sharp drop in $T_{c}$ from 15 K to $\sim$ 4 K is
observed at 10 GPa. Parallel room temperature synchrotron x-ray studies to 15
GPa suggest a structural transition near 10 GPa from rhombohedral to a higher
symmetry structure rather than the order-disorder transition suggested by
Gauzzi \textit{et al.} \cite{gauzzi2}.

\section{Experimental}

CaC$_{6}$ samples are prepared using the alloy method described by Emery
\textit{et al.} \cite{emery1}. A stainless steel tube is cleaned, baked at
900$^{\circ}$C in vacuum and loaded with lithium and calcium in the atomic
ratio 3:1. Natural Madagascar graphite flakes or highly oriented pyrolitic
graphite (HOPG) pieces (GE ZYA grade) are added to the Li-Ca alloy. The tube
is mechanically sealed in an argon atmosphere and placed inside a one-zone
furnace which is evacuated to 2x10$^{-7}$ Torr and subsequently filled with
argon gas. The furnace is then heated to 350$^{\circ}$C and the reaction takes
place for 10 days. After this time, the furnace is turned off allowing the
sample to slowly cool down to room temperature. The tube is opened inside an
argon-filled glove box and the samples, exhibiting a golden appearance,
removed by dissolving the alloy in ethylenediamine (Sigma Aldrich
$>$
99\%).

Fig. 1 shows the x-ray diffraction data for a typical CaC$_{6}$ flake sample
with the (00l) diffraction peaks obtained using a Rigaku x-ray diffractometer
with Cu-K$\alpha$ radiation and taken in a Bragg-Brentano geometry
\cite{emery2}. The pattern is consistent with the rhombohedra structure model
for CaC$_{6}$ of Emery \textit{et al}. \cite{emery1}. No lines corresponding
to hexagonal graphite are visible within our detection limits, showing the
bulk nature of the sample. From the diffraction data we find the lattice
parameters $a=4.33$ \AA \ and $c=13.57$ \AA , yielding a mass density of 2.53
g/cm$^{3}$.

For hydrostatic pressures to 0.6 GPa a He-gas compressor system from Harwood
Engineering was used in combination with a CuBe pressure cell from Unipress
with a 7 mm diameter bore. Using a primary/secondary compensated coil system
immediately surrounding the sample in the cell bore, ac susceptibility
measurements at 0.1 Oe rms and 1023 Hz can be carried out under pressure to
the same high accuracy as measurements at ambient pressure. A two-stage
Balzers closed-cycle refrigerator was used to cool the pressure cell to below
the superconducting transition temperature of CaC$_{6}$; all measurements were
carried out upon warming up slowly through the transition at the rate
$\sim0.06$ K/min. All susceptibility measurements were repeated at least once
to verify that the transition temperature at a given pressure was reproducible
to within 20 mK.

The membrane-driven diamond-anvil cell (DAC) \cite{schilling2} in this
experiment employed 1/6-carat type Ia diamond anvils with 0.5 mm culets and a
3 mm girdle. After the non-superconducting, non-magnetic NiMo-alloy gasket was
preindented from 380 $\mu$m to 80 $\mu$m, a 235 $\mu$m hole was spark-cut
through the center. Tiny ruby spheres \cite{chervin} are placed on or near the
sample to allow the pressure determination at a temperature near 20 K using
the revised ruby calibration of Chijioke \textit{et al.} \cite{chijioke}. The
ac susceptibility in the DAC is measured using two compensated
primary/secondary coil systems with an applied field of 3 Oe rms at 1023 Hz.
Further details of the DAC \cite{schilling2,hamlin} and He-gas compressor
\cite{tomita} techniques are given elsewhere.

Structural properties of CaC$_{6}$ as function of pressure up to 15 GPa were
studied in a DAC using angle-dispersive synchrotron x-ray diffraction
techniques at beamline 16ID-B of the High Pressure Collaborative Access Team
(HPCAT), Advanced Photon Source (APS). X-ray diffraction measurements were
carried out with a focused 33.714 keV monochromatic beam ($10\times10$ $\mu$m)
and a MAR345 image plate detector to record the diffracted x-rays. The
diffraction images were integrated using the software FIT2D \cite{hammersley}.
Silicon oil was used as pressure medium. Pressures were determined by the ruby
fluorescence method.

\section{Results of Experiment and Discussion}

\subsection{AC Susceptibility Measurements}

\subsubsection{He-Gas Compressor System}

In Figs.~2(a) and 2(b) the diamagnetic transition to superconductivity for
both CaC$_{6}$ samples is seen to shift to higher temperatures with increasing
hydrostatic pressure. The size of the superconducting shielding is more than
twice as large for the HOPG-graphite sample as for the natural graphite
sample. In Fig.~3 the superconducting transition temperature $T_{c}$ is
plotted versus pressure for both samples \cite{note1}. The measured $T_{c}%
(P)$-dependence is seen to be reversible and not depend on the temperature at
which the pressure is changed. Within experimental error the rate of increase,
$dT_{c}/dP\approx+0.40(1)$ K/GPa, is the same for both samples, but 5 - 10 \%
less than that reported previously in experiments where less hydrostatic
pressure media were used \cite{smith,kim,gauzzi}.

\subsubsection{Diamond Anvil Cell}

The results of the present ac susceptibility experiments on the HOPG graphite
CaC$_{6}$ sample in a DAC are shown in Fig.~4. Following the initial
pressurization to $\sim$ 5 GPa at 2 K, the pressure was changed in the
temperature range 100 - 150 K. The transition is seen to shift to higher
temperatures with pressure to 9.5 GPa, but then to suddenly fall to 8 K at
10.7 GPa and broaden. At 11.8 GPa $T_{c}$ lies near 4 K and decreases
moderately at higher pressures to 18.3 GPa. At higher pressures the transition
shifted to temperatures below our temperature window and did not reappear to
32 GPa.

In Fig.~5 this dependence of $T_{c}$ on pressure is shown explicitly and
compared to the previous results of Gauzzi \textit{et al.} \cite{gauzzi} to 18
GPa. In both measurements $T_{c}$ is seen to plummet downward rapidly at a
pressure near 10 GPa, indicating a possible first-order phase transition. This
possibility is supported by the sudden marked broadening of the transition at
10.7 GPa. The initial slope $dT_{c}/dP\approx+0.39(1)$ K/GPa to 9.5 GPa agrees
well with the results of our He-gas studies to 0.6 GPa in Fig.~3.

\subsection{Synchrotron X-Ray Diffraction Studies}

The x-ray diffraction patterns for CaC$_{6}$ (HOPG graphite) for increasing
pressure are shown in Fig.~6 and are indexed based on 9 to 12 diffraction
lines of an hexagonal unit cell consistent with the known rhombohedral
structure model of CaC$_{6}$ \cite{emery1}. The unit cell volume $V$ and
lattice parameters of this phase as a function of pressure are shown in
Figs.~7(a) and 7(b). The $P-V$ data are fit to a third-order Birch-Murnaghan
equation of state%
\begin{equation}
P=\frac{3}{2}\left[  \left(  \frac{V_{o}}{V}\right)  ^{7/3}-\left(
\frac{V_{o}}{V}\right)  ^{5/3}\right]  \left\{  1-\frac{3}{4}\left(
4-K_{o}^{\prime}\right)  \left[  \left(  \frac{V_{o}}{V}\right)
^{2/3}-1\right]  \right\}  ,
\end{equation}
where $V_{o}$ and $K_{o}$ are the volume and isothermal bulk moduli,
respectively, at 1 bar, and $K_{o}^{\prime}$ is the pressure derivative of
$K_{o}$. By fixing $K_{o}^{\prime}$ = 4, a least-squares fit yields $K_{o}$ =
119(3) GPa, and $V_{o}$ = 220.5(3). As expected for most layered structures,
the compression of the low-pressure CaC$_{6}$ phase is anisotropic where the
$c$ axis is more than 10 times more compressible than the $a$ axis, with the
result that the $c/a$ ratio decreases appreciably with pressure (see inset in
Fig.~7(b)). From the data in Fig.~7(b) we find $\varkappa_{a}\equiv
-a^{-1}da/dP\simeq0.00039$ GPa$^{-1}$ and $\varkappa_{c}\equiv-c^{-1}%
dc/dP\simeq0.0069$ GPa$^{-1}$.

The low-pressure superconducting CaC$_{6}$ phase is stable to 9.0 GPa (Figs.~6
and 8). Between 9 and 10.5 GPa the x-ray diffraction pattern displays evident
changes, including the disappearance of some of the diffraction lines of the
low-pressure CaC$_{6}$ phase and a significant intensity decrease in the
region of the previous (113) diffraction line, the strongest of the
low-pressure CaC$_{6}$ structure, as well as the appearance of new diffraction
lines. These observations suggest a structural phase transformation, likely to
a higher symmetry phase.

In summary, the dependence of the superconducting transition temperature of
CaC$_{6}$ on nearly hydrostatic pressure has been studied to 32 GPa. Following
an increase from 11 K to 15 K under 9.5 GPa pressure, $T_{c}$ abruptly drops
to 4 K at 11.8 GPa. Parallel synchrotron x-ray studies reveal that this sudden
drop in $T_{c}$ results from a structural phase transition from rhombohedral
to a phase of higher symmetry.\vspace{0.3cm}

\noindent Acknowledgments. Thanks are due V. G. Tissen for supplying the
NiMo-alloy gaskets used in the DAC experiments. The authors at Washington
University gratefully acknowledge research support by the National Science
Foundation through Grant No. DMR-0703896. HPCAT is supported by DOE-BES,
DOE-NNSA (CDC), NSF, and the W. M. Keck Foundation. APS is supported by
DOE-BES, under Contract No. DE-AC02-06CH11357.

\begin{center}
\bigskip{\LARGE Figure Captions}
\end{center}

\noindent\textbf{Fig. 1. \ }X-ray (Cu K$\alpha$) diffraction pattern for
CaC$_{6}$ sample prepared from natural graphite.\bigskip\ 

\noindent\textbf{Fig. 2. \ }(color online) Real part of the ac susceptibility
versus temperature at different pressures to $\sim$ 0.6 GPa in the He-gas cell
for CaC$_{6}$ made from (a) natural graphite or (b) HOPG graphite. \bigskip

\noindent\textbf{Fig. 3. \ }(color online) Superconducting transition
temperature versus pressure to $\sim$ 0.6 GPa for CaC$_{6}$ using data from
Figs.~2(a) and 2(b). $T_{c}$ is determined from the transition midpoint (see
Ref. \cite{note1}). Numbers give order of measurement. Pressure is changed at
room temperature for solid symbols, at 50 K for open symbols.\bigskip

\noindent\textbf{Fig.\ 4.\ \ }Real part of the ac susceptibility signal at
different pressures in the DAC to 18.3 GPa for CaC$_{6}$ made from HOPG
graphite. Data taken with increasing pressure. \bigskip

\noindent\textbf{Fig. 5. \ }Superconducting transition temperature versus
pressure to 32 GPa for CaC$_{6}$ using data from Fig.~4. $T_{c}$ is determined
from the transition midpoint. Solid line is guide to eye; dashed line
reproduces data from Ref. \cite{gauzzi}. Above 20 GPa $T_{c}$ lies below 2 K.
Error bars give 20\% - 80\% transition width.\bigskip

\noindent\textbf{Fig. 6. \ }Synchrotron x-ray patterns for HOPG CaC$_{6}$
under pressure to 15 GPa. A structural phase transition appears to occur for
pressures above 9 GPa. Diffraction peaks in the 2-theta angle from 5 to 7
degrees are not used for unit cell calculations due to the overlap of large
single crystal spots in this region.\bigskip

\noindent\textbf{Fig. 7. \ }Unit cell volume and normalized hexagonal lattice
parameters $a$ and $c$ as a function of pressure to 9 GPa. Inset shows the
$c/a$ ratio versus pressure.\bigskip

\noindent\textbf{Fig. 8. \ }$d$-space versus pressure for CaC$_{6}$ to 15
GPa.\newpage%

{\parbox[b]{5.467in}{\begin{center}
\includegraphics[
natheight=5.480300in,
natwidth=5.920300in,
height=5.0635in,
width=5.467in
]%
{graphs/Fig12.jpg}%
\\
Figure 1
\end{center}}}%
%

{\parbox[b]{4.6318in}{\begin{center}
\includegraphics[
natheight=7.760000in,
natwidth=10.137800in,
height=3.5517in,
width=4.6318in
]%
{graphs/Fig1a.wmf}%
\\
Figure 2(a)
\end{center}}}%
%

{\parbox[b]{4.6733in}{\begin{center}
\includegraphics[
natheight=7.760000in,
natwidth=10.137800in,
height=3.5832in,
width=4.6733in
]%
{graphs/Fig1b.wmf}%
\\
Figure 2(b)
\end{center}}}%
%

{\parbox[b]{4.7214in}{\begin{center}
\includegraphics[
natheight=13.292600in,
natwidth=10.172600in,
height=6.1569in,
width=4.7214in
]%
{graphs/Fig2.wmf}%
\\
Figure 3
\end{center}}}%
%

{\parbox[b]{4.6052in}{\begin{center}
\includegraphics[
natheight=7.833100in,
natwidth=10.176000in,
height=3.5508in,
width=4.6052in
]%
{graphs/Fig3.wmf}%
\\
Figure 4
\end{center}}}%
%

{\parbox[b]{4.6301in}{\begin{center}
\includegraphics[
natheight=7.833100in,
natwidth=10.176000in,
height=3.5708in,
width=4.6301in
]%
{graphs/Fig4.wmf}%
\\
Figure 5
\end{center}}}%
%

{\parbox[b]{5.8912in}{\begin{center}
\includegraphics[
natheight=4.214200in,
natwidth=3.123300in,
height=7.9327in,
width=5.8912in
]%
{graphs/Fig7.wmf}%
\\
Figure 6
\end{center}}}%
%

{\parbox[b]{5.0668in}{\begin{center}
\includegraphics[
natheight=4.195100in,
natwidth=3.251100in,
height=6.5247in,
width=5.0668in
]%
{graphs/Unit_Cell.wmf}%
\\
Figure 7
\end{center}}}%
%

{\parbox[b]{4.5737in}{\begin{center}
\includegraphics[
natheight=4.483200in,
natwidth=3.257800in,
height=6.2789in,
width=4.5737in
]%
{graphs/d-space.wmf}%
\\
Figure 8
\end{center}}}%


\begin{thebibliography}{99}
\bibitem {note1}At ambient pressure (pt. 1 in Fig. 3) $T_{c}$ is defined from
the midpoint of the superconducting transition. $T_{c}$ at high pressure is
most accurately determined from the shift of the entire transition curve
relative to that at ambient pressure.\newpage
\end{thebibliography}
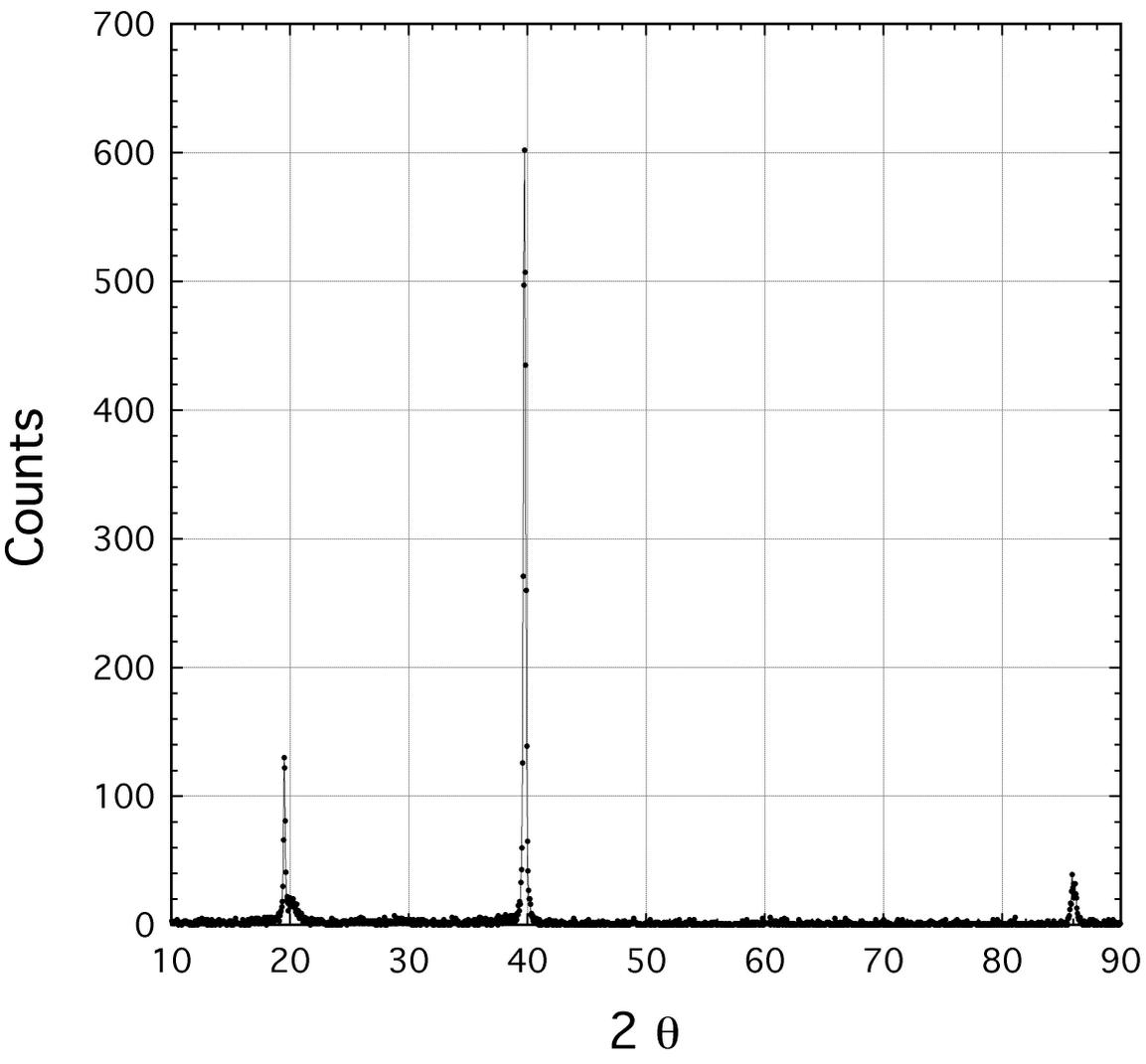
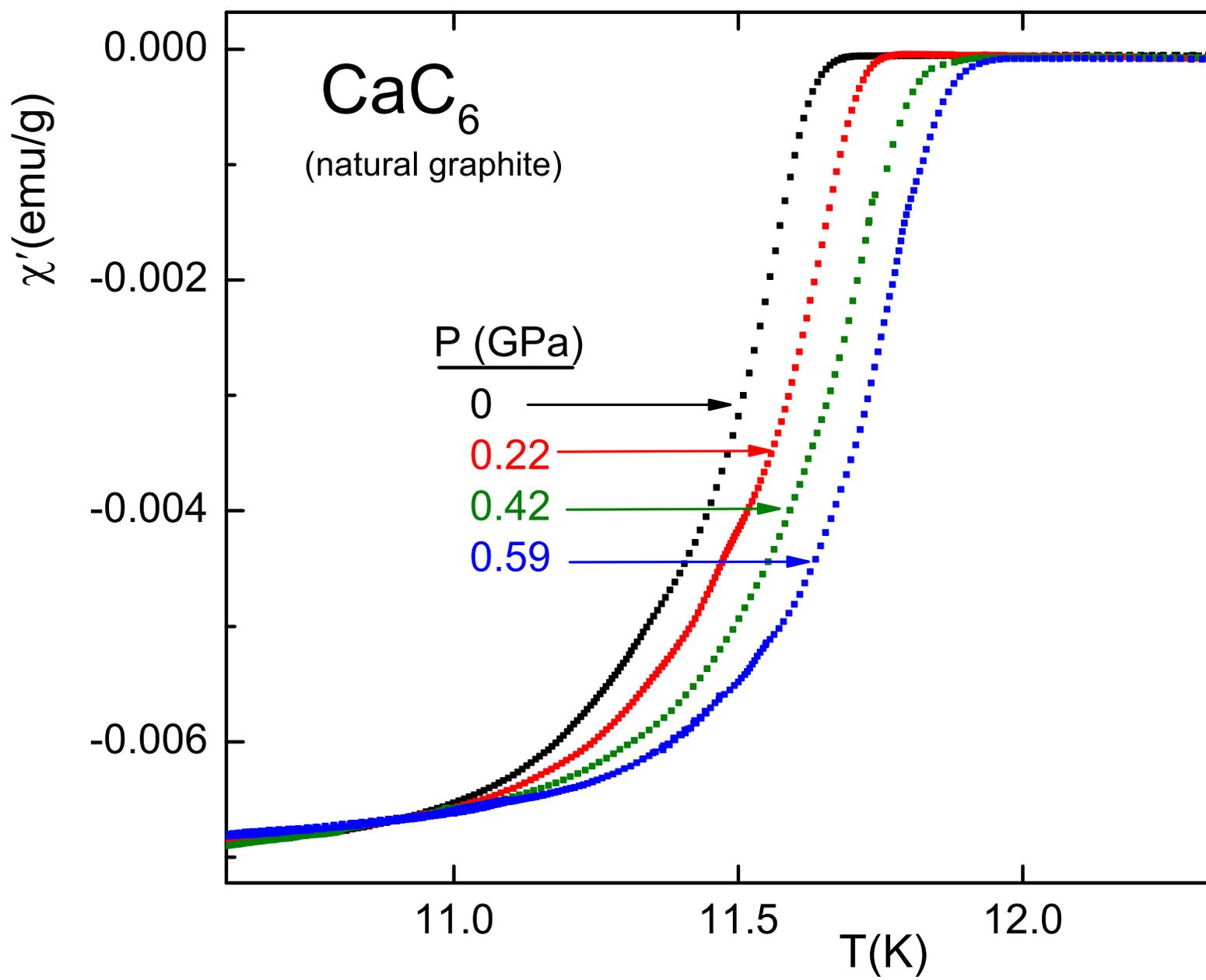
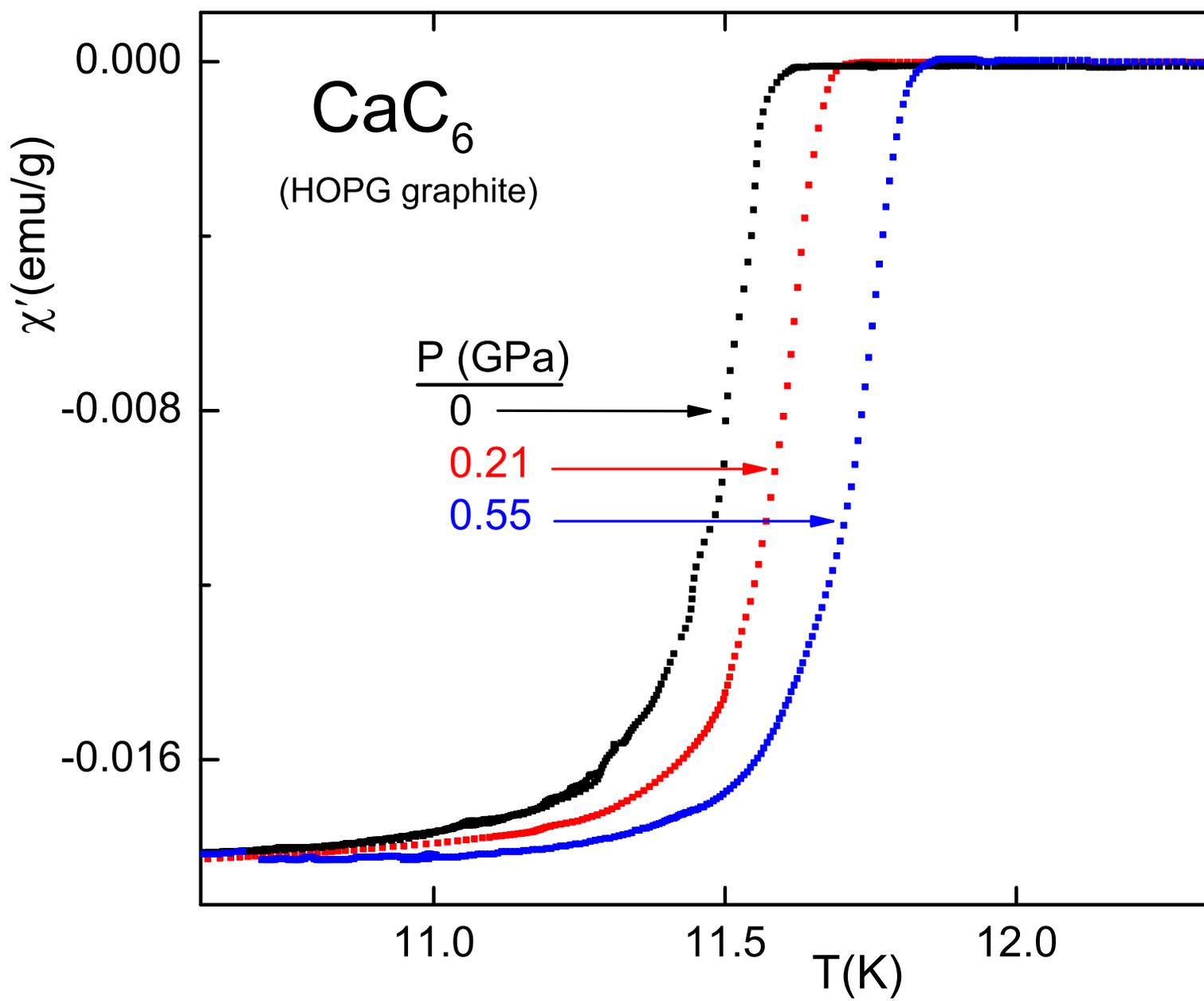
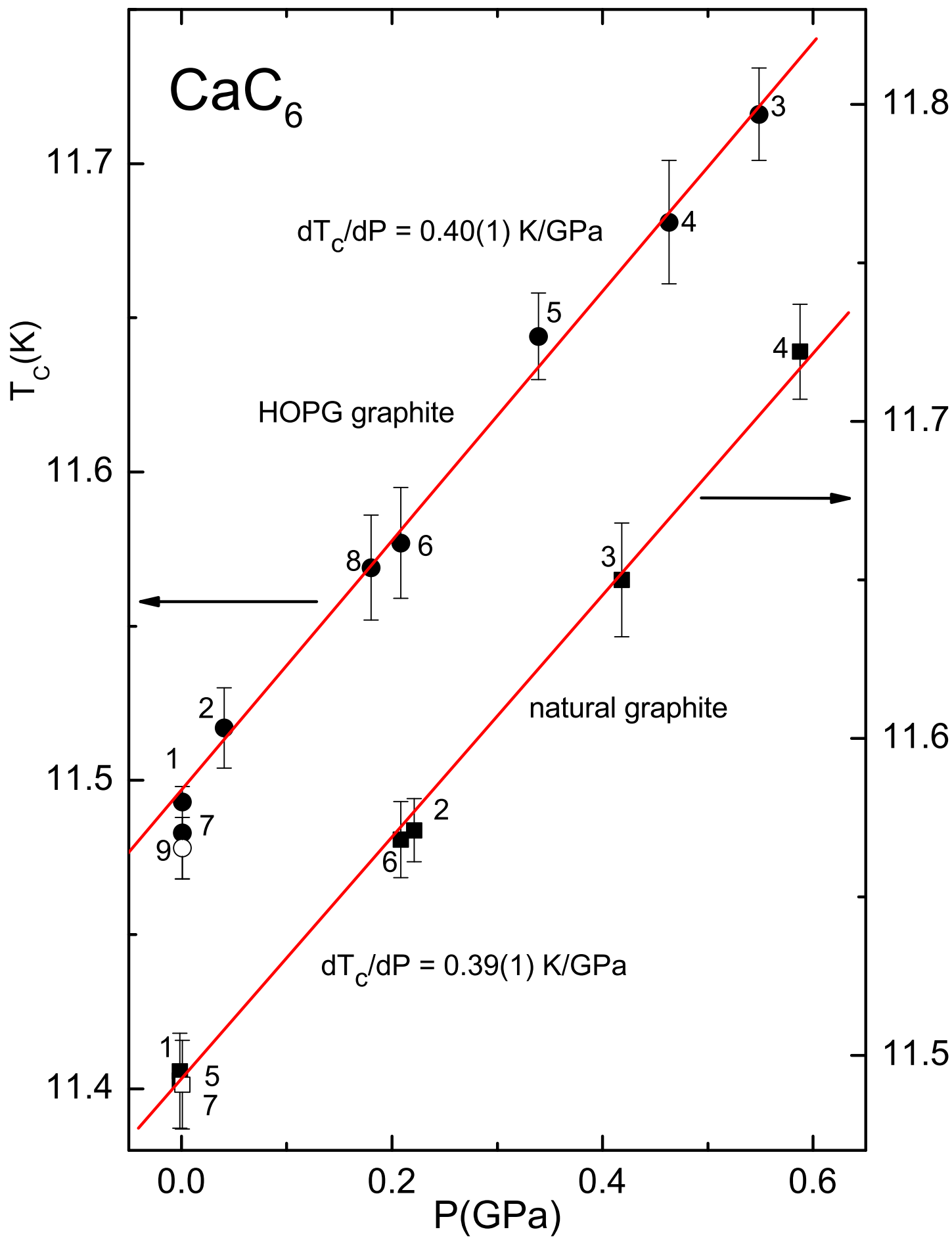
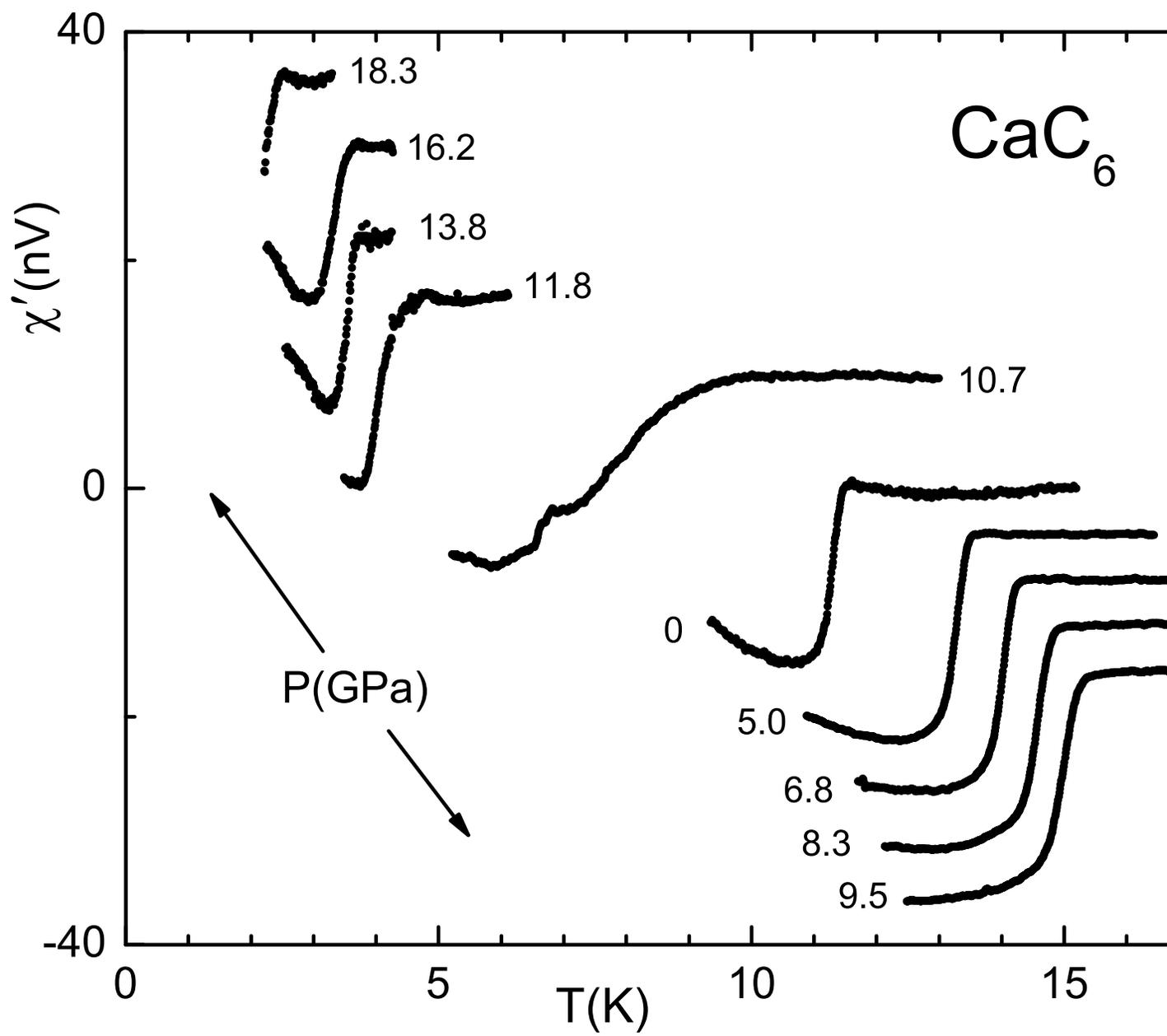
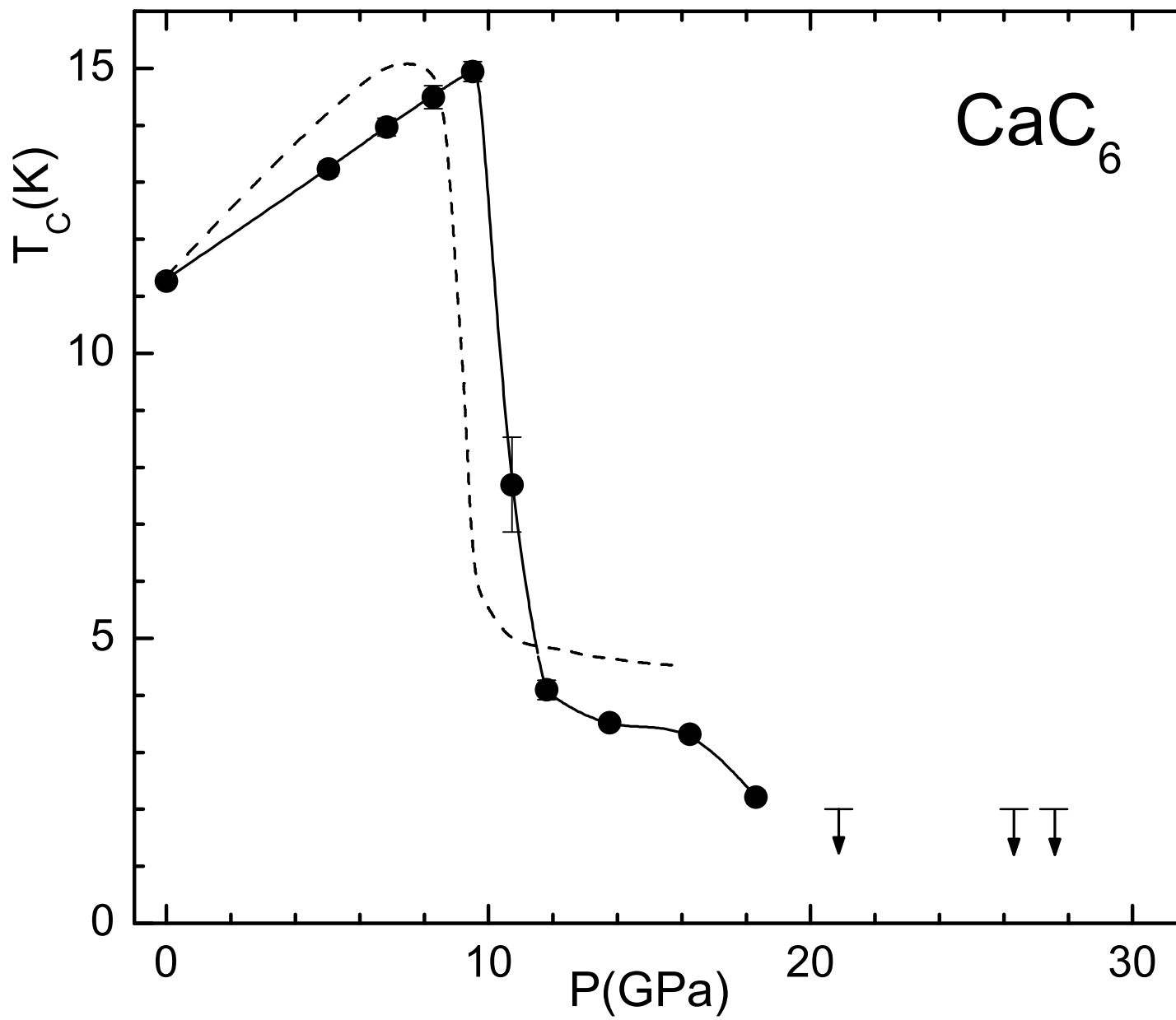
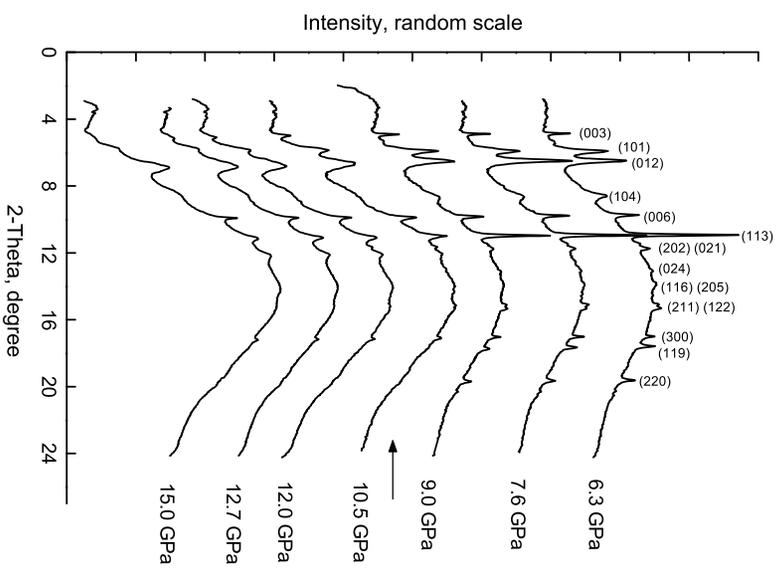
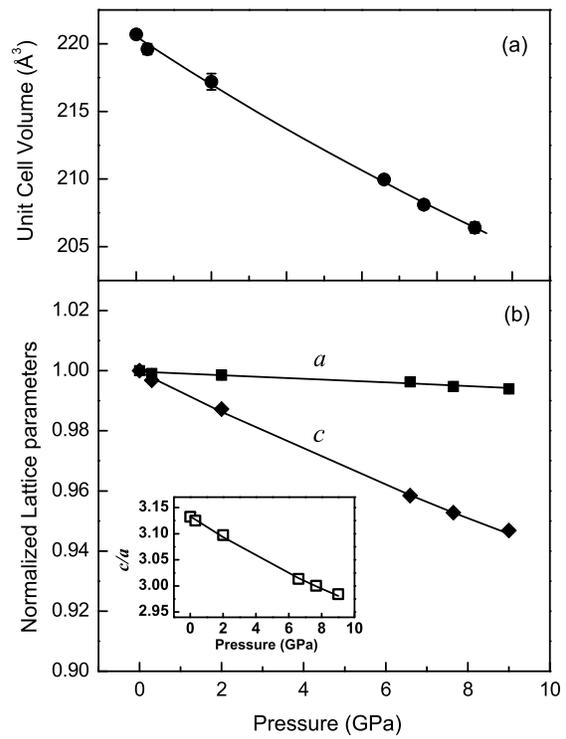
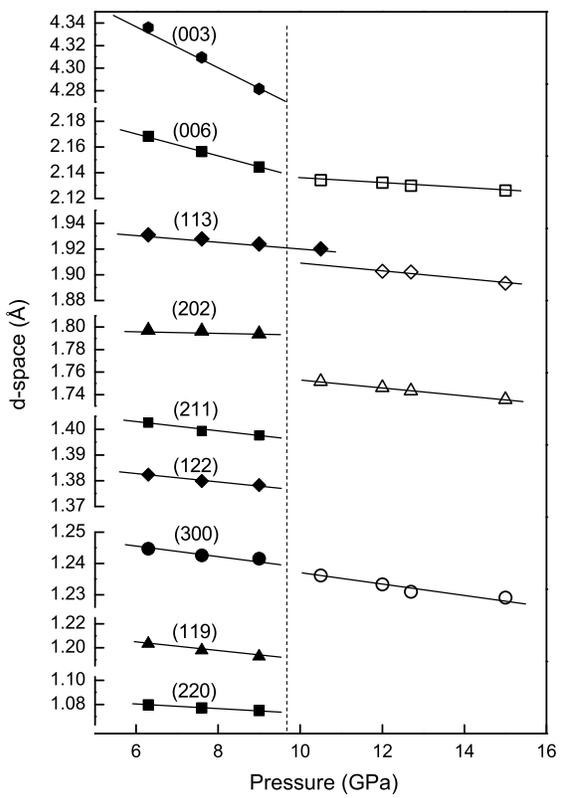
\end{document}